\documentclass[12pt]{article}
\usepackage{geometry}             
\usepackage{graphicx}
\usepackage{amssymb}
\usepackage{amsmath}
\usepackage{subfigure}
\usepackage{cite}

\usepackage[hyperindex=true,
          pdfstartview=FitH,
          bookmarksnumbered=true,
          bookmarksopen=true,
          citecolor=blue,
          linkcolor=blue,
          colorlinks=true,
          unicode]{hyperref}

\begin{document}

\title{C-metric solution for conformal gravity with a conformally coupled scalar field}
\author{Kun Meng${}^{1}$ and Liu Zhao${}^{2}$\\
${}^{1}$ School of Science, Tianjin Polytechnic University, \\
Tianjin 300387, China\\
${}^{2}$ School of Physics, Nankai University, Tianjin 300071, China\\
emails: \href{mailto:mengkun@tjpu.edu.cn}{\it mengkun@tjpu.edu.cn}
and \href{mailto:lzhao@nankai.edu.cn}{\it lzhao@nankai.edu.cn}}
\date{}                             
\maketitle

\begin{abstract}

The C-metric solution of conformal gravity with a conformally coupled scalar field is presented.
The solution belongs to the class of Petrov type D spacetimes and is conformal to the standard
AdS C-metric appeared in vacuum Einstein gravity.
For all parameter ranges, we identify some of the physically interesting static regions and the
corresponding coordinate ranges. The solution may contain a black hole event horizon,
an acceleration horizon, either of which may be cut by the conformal infinity or be
hidden behind the conformal infinity. Since the model is conformally invariant,
we also discussed the possible effects of the conformal gauge choices on the
structure of the spacetime.
\end{abstract}

\section{Introduction}

Conformal gravity is a kind of higher curvature gravity which is invariant under
conformal transformations. In the simplest form in four spacetime dimensions, its action is
consisted of the square of Weyl curvature tensor. Because of the conformal invariance, this
model is only sensitive to angles but not to distances. This feels counter intuition since
everyone knows that gravity should decrease as the distance increases.
On the level of linear perturbations, conformal gravity suffers from the existence of
ghost degrees of freedom, which implies vacuum instability. For these reasons,
conformal gravity may well be thought of as an unphysical model of gravity.
However, as Maldacena pointed out \cite{Maldacena}, the on-shell action of conformal gravity is
identical to that of Einstein gravity in an (A)dS background, and under proper boundary
conditions (Neumann boundary conditions), the contribution from the ghost degrees of freedom
can be removed. Conversely, the boundary anomaly of five dimensional Einstein gravity yields
the action of four dimensional conformal gravity \cite{Maldacena}, and also, the four
dimensional Einstein gravity with a negative cosmological constant regularized with a
topological Gauss-Bonnet term can be reduced on-shell to that of the conformal gravity
\cite{Olev}.
Therefore, it is reasonable to take conformal gravity more seriously than just another toy model
of gravity.

Maldacena's arguments on the equivalence between conformal gravity and Einstein gravity
are subjects to several limitations. It works only in the classical (or tree-level) regime,
only for backgrounds which are Einstein manifolds obeying the Neumann boundary conditions
and only in the vacuum sector. When any of these limitations is broken, it is unclear whether we can
still treat conformal gravity as an equivalent model of Einstein gravity. Take the matter
contribution for instance. When matter sources are present, it is not clear {\em a priori}
whether Einstein manifolds are solutions to the model and whether the boundary conditions
proposed by Maldacena can still be applied. For generic matter sources, the answer to these
problems might be ``no'', because the boundary conditions actually played an important role
in breaking the conformal invariance in some way, but the matter source may have already
broken the conformal invariance which leaves no room for breaking the conformal invariance in
other ways. But there is an interesting class of matter sources, i.e. conformally coupled
matter sources, which keeps the conformal invariance of the model. Conformal gravity with
conformally coupled matter sources might still provide room for arguments which are analogous
to what Maldacena has made in the vacuum sector.

However, an obvious obstacle will emerge when one tries to mimic Maldacena's arguments
in the cases with conformally coupled sources, i.e. sufficient message on the solutions to
such cases is a necessary input. Unlike the case of vacuum conformal gravity which is studied
extensively in the literature (and it is known that any spacetime that is conformal to an Einstein
manifold is a solution \cite{Maldacena}),
known solutions to conformal gravity with conformally coupled
sources are relatively rare. An outstanding exception is conformal gravity with minimally coupled
electromagnetic field. For this particular type of conformal invariant matter source,
a number of exact solutions with different symmetries have been found \cite{MannheimKazanas,Said}.
An extension to the case with an additional SU(2) Yang-Mills source has also been
found to possess exact AdS black hole solutions \cite{HongLu2}. Besides these particular
cases, no other solutions to conformal gravity with conformally coupled matter sources have been
known, at least to our knowledge.

In this paper, we aim to present exact solutions to conformal gravity with a conformally coupled
scalar field (CGCCS model for short). The type of solution which we will present is
quite similar to the C-metric which has been known for a long time
\cite{Levi-Civita,EhlersKundt,Weyl1917} in the context of pure Einstein gravity.
The reason that we pay our attention
to the C-metric-like solutions is because such solutions capture almost every aspects of
classical relativistic spacetimes and therefore can be used as a nontrivial theoretical
laboratory for studying relativistic spacetimes. For instance, they contain black holes --
actually two black holes accelerating apart -- and hence also acceleration horizons
\cite{KinnersleyWalker,Bonnor,GriffithsPodolsky}. They can be easily generalized to
bear cosmological constants \cite{DisaLemos2002,DisaLemos2003} and/or electromagnetic charge
\cite{KinnersleyWalker}. For Einstein gravity conformally coupled with a scalar field, the
C-metric solution was also found \cite{Charmousis0906,Anabalon0907}. The rotating form of the
C-metric is known as Plebanski-Demianski metric \cite{PB}. Finally, the black ring solution
\cite{blackring} which appears in five dimensional Einstein gravity contains a Wick-rotated
version of the C-metric as
a building block. We hope that the rich structure of the C-metric may also be useful in
exploring the possible connections between conformal gravity and Einstein gravity,
or in revealing the differences of conformal gravity from Einstein gravity.
As a note in passing, let us remark that, even in the vacuum sector, the C-metric solution in
conformal gravity has not been explored (though expected to exist). This is in contrast to
the situations of non-accelerating black hole solutions which have been studied extensively
\cite{Said,HongLu2,Riegert,HongLu1,HongLu3,HongLu4}.

The paper is organized as follows. In Section \ref{section2}, we give the action,
field equations and the C-metric solution of of the CGCCS model. It is shown that
the metric we get represents a spacetime which has a non-constant scalar curvature, but
is of Petrov class D, just like the standard C-metric in Einstein gravity.
Then, in Section \ref{sec3}, we study coordinate ranges and horizon structures of the solution,
with emphasis on the determination of the boundaries of the physically interesting
static regions (by which we mean the static region outside a black hole event horizon).
This section is subdivided into three subsections according to the different values
of the scalar self-coupling $\lambda$ and another parameter $e_2$ entering the explicit
solution of the scalar field. In Section \ref{sec4} we consider two other
conformal gauges and show that the solution we get is conformal to the standard
AdS C-metric appeared in vacuum Einstein gravity. Finally, in Section {section4},
we give some concluding remarks.

\section{The model and the solution\label{section2}}

Up to a boundary counter term which is irrelevant to our study in this paper, the action of the
CGCCS model that we shall study is given as follows:
\begin{align}
I=\int \mathrm{d}^4x \sqrt{-g}\left[\frac{1}{2}\alpha C^{\mu\nu\rho\sigma}C_{\mu\nu\rho\sigma}
-\left(\frac{1}{2}g^{\mu\nu}\partial_\mu \Phi\partial_\nu \Phi
+\frac{1}{12} \Phi^2 R+\lambda\Phi^4\right)\right]\label{action},
\end{align}
where $C_{a b c d}$ and $R$ are respectively the Weyl tensor and Ricci scalar associated with
the spacetime metric $g_{ab}$, $\alpha$ and $\lambda$ are dimensionless coupling constants.
Note that the coefficient in front of the $\Phi^2R$ term is fixed by the requirement of
conformal invariance under the transformations
\begin{align}
g_{\mu\nu}\rightarrow \Omega^2(x)g_{\mu\nu},\quad
\Phi\rightarrow\Omega^{-1}(x)\Phi,
\label{conft}
\end{align}
so, there can be no other free parameters in the model. Note also that, by a field redefinition
$\Phi\to \alpha^{1/2}\Phi$ together with a rescaling of parameter $\lambda\to \alpha^{-1}
\lambda$, the parameter $\alpha$ becomes an overall factor in the action so that it plays no
role on the classical level. Therefore, we can set $\alpha=1$ without loss of generality.
In order that the scalar self-interacting
potential to be bounded from below, it is necessary to require $\lambda\geq 0$.

The field equations that follow from variations with respect to the metric $g_{ab}$ and to the
scalar field $\Phi$ are given respectively as
\begin{align}
B_{\mu\nu}&= -T^{(\Phi)}_{\mu\nu},\\
\Box\Phi&=\frac{1}{6}R\Phi+4\lambda\Phi^3,
\end{align}
where
\begin{align}
B_{\mu\nu} = 2\nabla^\rho \nabla^\sigma C_{\mu \rho\sigma \nu}
+R^{\rho\sigma}C_{\mu \rho\sigma \nu}
\end{align}
is the Bach tensor and
\begin{align}
&T^{(\Phi)}_{\mu\nu}=\partial_\mu\Phi\partial_\nu\Phi
-\frac{1}{2}g_{\mu\nu}\partial^\rho\Phi\partial_\rho\Phi
-\lambda g_{\mu\nu}\Phi^4+\frac{1}{6}\left(g_{\mu\nu}\Box
-\nabla_\mu\nabla_\nu+G_{\mu\nu}\right)\Phi^2
\end{align}
is the stress-energy tensor for the scalar field $\Phi$.

In the spacetime coordinates $x^\mu=(t,y,x,\sigma)$,
we take the metric ansatz
\begin{align}
\mathrm{d}s^2=\frac{1}{A^2(x-y)^2}\left(-F(y)\mathrm{d}t^2+\frac{\mathrm{d}y^2}{F(y)}
+\frac{\mathrm{d}x^2}{G(x)}+G(x)\mathrm{d}\sigma^2\right),\label{ansatz}
\end{align}
and meanwhile assume the scalar field to take the form
\begin{align}
\Phi(x,y)=\frac{e_1(x-y)}{x+y-e_2}, \label{Phi}
\end{align}
where $A, e_1, e_2$ are all constants.

By brute force using {\it Maple}, we arrive at the following solution:
\begin{align}
G(x)&=\frac{1}{6}C_1x^3+\frac{1}{2}C_2x^2+C_3x+C_4,\nonumber\\
F(y)&=\frac{1}{6}C_1y^3-\frac{1}{2}(C_1e_2+C_2)y^2+\left(\frac{1}{2}C_1e_2^2+C_2e_2+C_3\right)y
\nonumber\\
&\quad -\left(\frac{1}{6}C_1e_2^3+\frac{1}{2}C_2e_2^2+C_3 e_2+C_4
-\frac{2e_1^2\lambda}{A^2}\right),
\label{FnG}
\end{align}
where $C_i (i=1,\cdots,4)$ are integration constants. Not all of these constants are
necessarily important. We can make use of the coordinate transformations to fix  some of these
constants and reduce the solution to a simpler form. To be more specific, we can use the
following 3-parameter coordinate
transformations
\begin{align}
t\to b\, t,\quad y\to a \,b\, y-c, \quad x \to a\, b\, x -c, \quad \sigma \to b\, \sigma
\label{abctrans}
\end{align}
to fix or constrain three of the four integration constants $C_i$ at the sacrifice of
a rescaling of the constant $A$ and a shift and rescaling of the constant $e_2$. The constant
$e_1$ is not affected by such operations. In the following, we shall take the liberty of the
above coordinate degrees of freedom to set the integration constants $C_i$ to the specific values
\begin{align}
&C_1=-12mA,\quad C_2=-2,\nonumber\\
&C_3=2mA, \qquad\; C_4=1,\label{Constants}
\end{align}
where $m$ is the only residual free integration constant, which may be chosen to be always
positive thanks to the transformation rules \eqref{abctrans}.
Notice that we still denote the rescaled constant $A$ and the shift-rescaled $e_2$ by the same
symbols. Under the above choice of integration constants, the metric functions $G(x)$ and
$F(y)$ can be arranged in the form
\begin{align}
& G(x) =(1-x)(1+x)(1+2mAx), \label{Gx}\\
& F(y) = -[1-(e_2-y)][1+(e_2-y)][1+2mA(e_2-y)]+\frac{2e_1^2\lambda}{A^2}. \label{Fy}
\end{align}
It is straightforward to observe that
\begin{align}
F(\xi)=-G(e_2-\xi)+\frac{2e_1^2\lambda}{A^2}. \label{FG}
\end{align}

The metric \eqref{ansatz} with $G(x)$ and $F(y)$ given respectively by \eqref{Gx} and \eqref{Fy}
looks extremely similar to the C-metric solution of vacuum Einstein gravity, therefore
we call this solution the C-metric solution for the CGCCS model. However, there are
some significant differences from the standard C-metric in Einstein gravity.
Among these, let us point out two major differences:
\begin{itemize}
\item Unlike the vacuum C-metric in Einstein gravity, our solution is sourced, sensitive to the
scalar field (which can be seen from the appearance of the constants $e_1,e_2$ and $\lambda$ in
\eqref{Gx} and \eqref{Fy});
\item Our solution is not a constant curvature spacetime. By straightforward calculations
and some elementary algebraic manipulations, the Ricci scalar of our solution can be written
in the following form:
\begin{align*}
R&=-24 e_1^{2}\lambda-2\,{A}^{2} \big\{ (x+y)^2+2xy \\
&\quad+6(x+y-e_2)[(2\,mA-e_2)
+mA(e_2(x+y-e_2)-2xy-e_2^2)]
\big\}. 
\end{align*}
So, it will be interesting to ask whether our solution is conformal to an
Einstein manifold. We postpone the answer of this question to Section \ref{sec4}.
\end{itemize}
In spite of the differences mentioned above, there is a crucial similarity between our solution
and the standard C-metric. By choosing the null Newman-Pensrose tetrads
\begin{align*}
&l^\mu = \frac{1}{\sqrt{2}}(T^\mu + Y^\mu),&&
n^\mu =  \frac{1}{\sqrt{2}}(T^\mu - Y^\mu),\\
&m^\mu =\frac{1}{\sqrt{2}}(X^\mu +i S^\mu),&&
\bar m^\mu =\frac{1}{\sqrt{2}}(X^\mu -i S^\mu),
\end{align*}
where
\begin{align*}
&T^\mu = \left(\frac{A(x-y)}{\sqrt{F(y)}},0,0,0\right),&&
Y^\mu =\left(0,A(x-y)\sqrt{F(y)},0,0\right),\\
&X^\mu = \left(0,0,A(x-y)\sqrt{G(x)},0\right),&&
S^\mu=\left(0,0,0,\frac{A(x-y)}{\sqrt{G(x)}}\right),
\end{align*}
we find that the only nonvanishing projection of the Weyl curvature of our spacetime on
the Newman-Penrose null tetrads is
\[
\Psi_2 = C_{\mu\nu\rho\sigma} n^\mu m^\nu \bar m^\rho l^\sigma = -mA^3(x-y)^2(x+y-e_2).
\]
Therefore, our solution is of Petrov type D,
just like the standard C-metric in vacuum Einstein gravity.

\section{Coordinate ranges and horizon structures} \label{sec3}

In this section, we would like to analyze some aspects of the structure of the spacetime solution
given by eqs. \eqref{ansatz}, \eqref{Gx} and \eqref{Fy}. We will be particularly interested in the
understandings about the ranges of the coordinates, the horizon structures and the interpretation
of the constant parameters. Essentially we will be following the lines of \cite{GriffithsPodolsky}
and that of \cite{Teo1}.

According to \eqref{Fy}, the root structure of the function $F(y)$ is different for the cases
$\lambda=0$ and $\lambda\neq0$. In the former case, the function $F(y)$ is already factorized,
from which we can read off explicitly the three roots, which are all independent of the constant
$e_1$. In the latter case, however, the function $F(y)$ is not factorized, and it is more
difficult to find its root structure. So, we will proceed differently for these two cases.

\subsection{The case $\lambda=0$\label{section2.1}}

When $\lambda=0$, $F(y)$ is explicitly factorized, so it is easy get the roots of $F(y)$:
\begin{align}
y_1=e_2-1,\quad y_2 = e_2+1,\quad y_3=e_2+ \omega^{-1}.\label{y1y2y3}
\end{align}
Here we have introduced the shorthand notation
\[
\omega\equiv 2mA
\]
because this expression will appear repeatedly in the forth coming discussions.
To fix the order of the three roots, we assume $0<\omega<1$ which can always be achieved using the
transformations given in \eqref{abctrans}. Under this assumption, we have
\[
y_1<y_2<y_3,
\]
regardless of what value the constant $e_2$ takes (provided it is real).

\begin{figure}[h]
\begin{center}
\includegraphics[width=.32\textwidth]{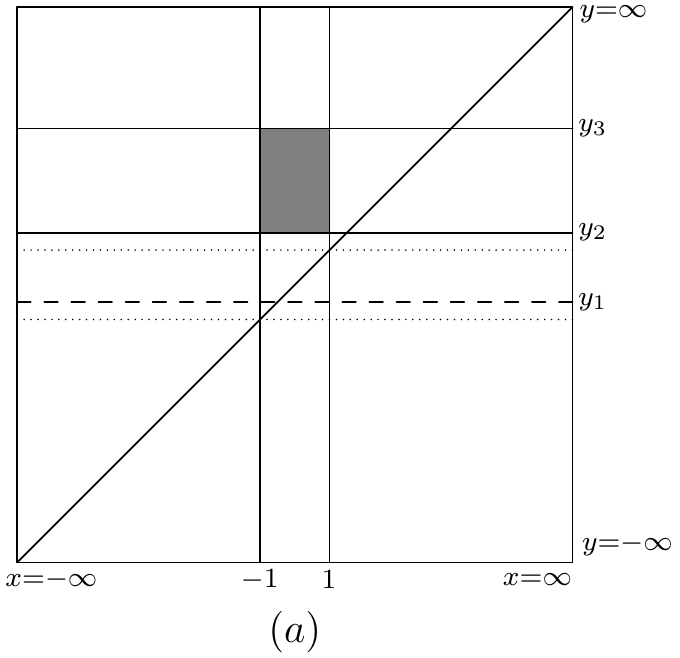}
\includegraphics[width=.32\textwidth]{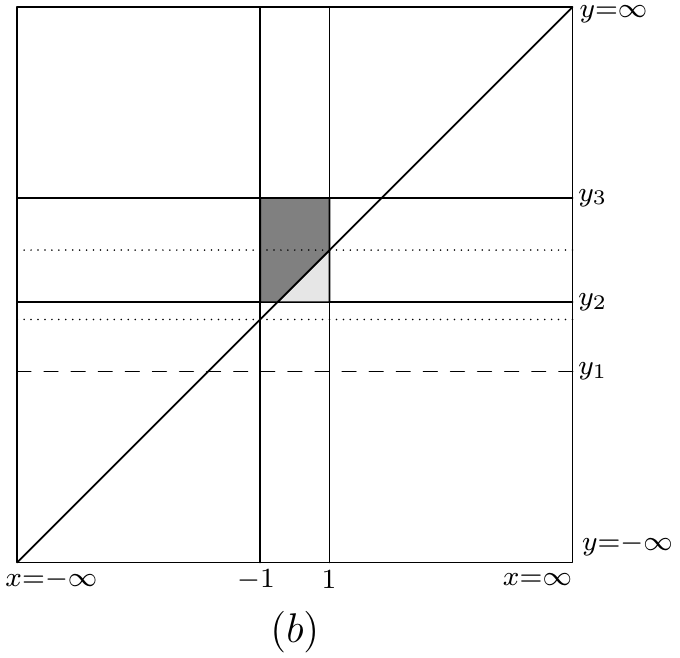}
\includegraphics[width=.32\textwidth]{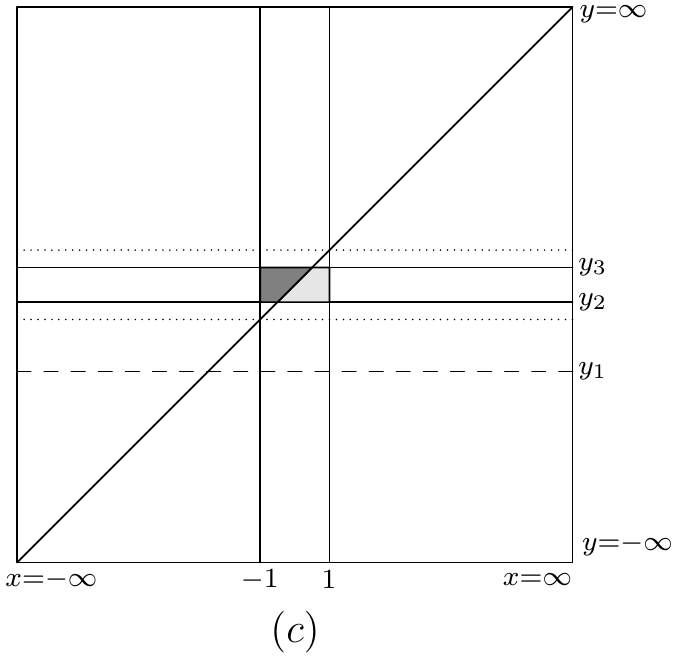}
\end{center}
\begin{center}
\includegraphics[width=.32\textwidth]{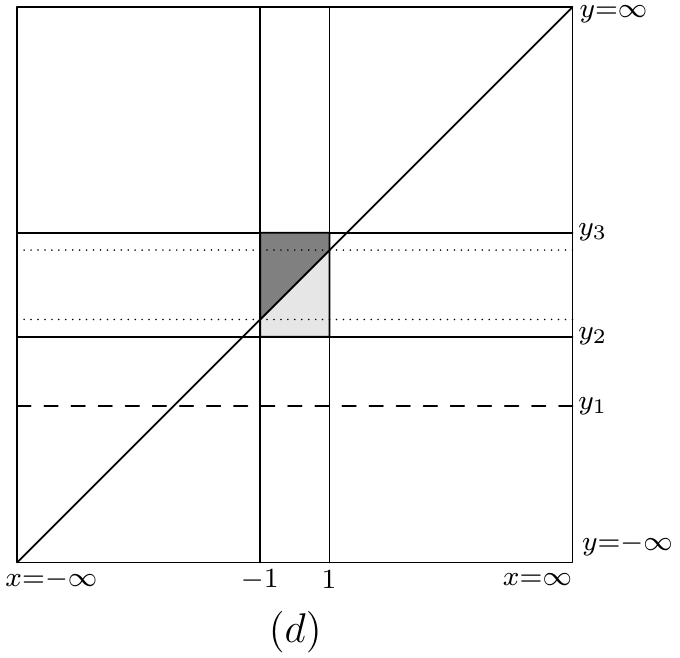}
\includegraphics[width=.32\textwidth]{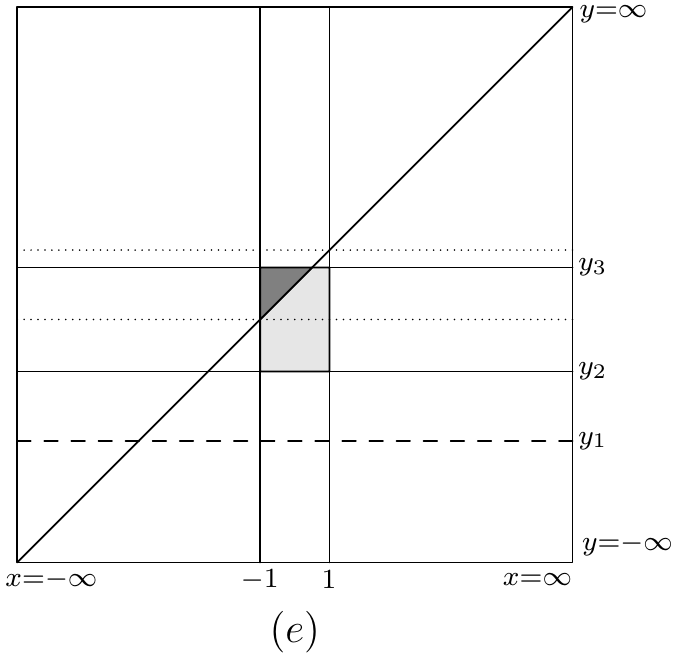}
\includegraphics[width=.32\textwidth]{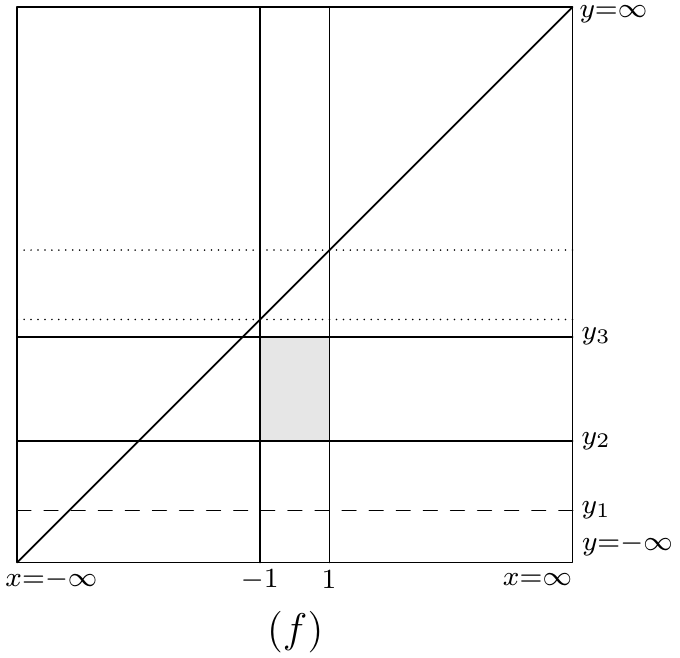}
\end{center}
\caption{Physically interesting static regions: dotted lines correspond to $y=\pm1$.
($a$) $e_2>0$; ($b$) $-2<e_2<0$ and
$\omega^{-1}>1-e_2$; ($c$) $-2<e_2<0$ and $\omega^{-1}<1-e_2$;
($d$) $e_2<-2$ and $\omega^{-1}>1-e_2$; ($e$) $e_2<-2$ and $-e_2-1<\omega^{-1}<1-e_2$;
($f$) $e_2<-2$ and $0<\omega^{-1}<-e_2-1$.}
\label{fig1}
\end{figure}

\begin{figure}[h]
\begin{center}
\includegraphics[width=.32\textwidth]{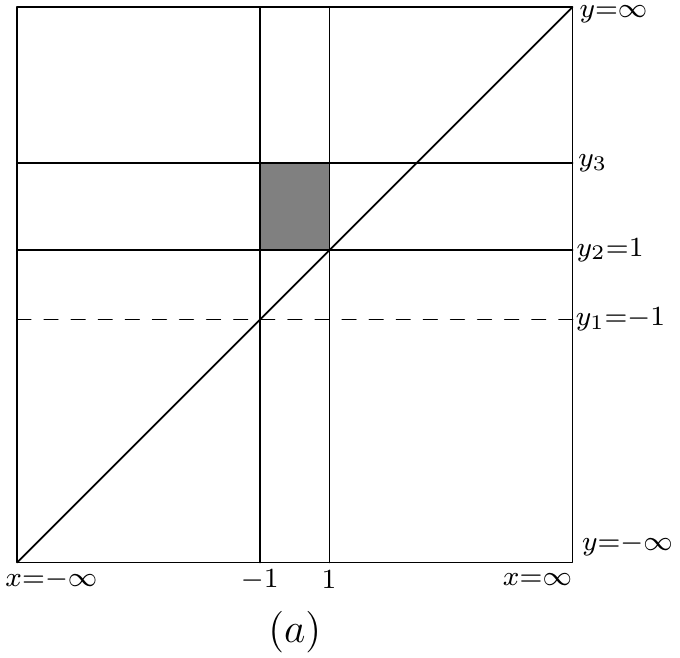}
\includegraphics[width=.32\textwidth]{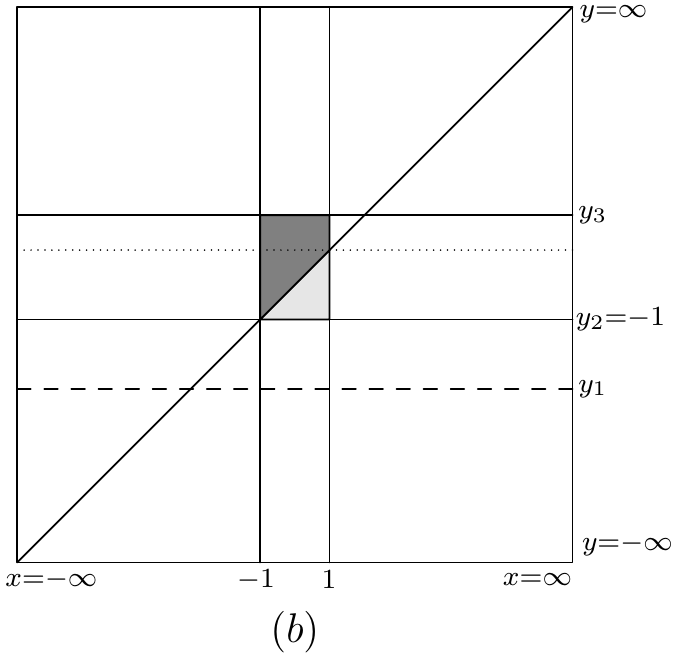}
\includegraphics[width=.32\textwidth]{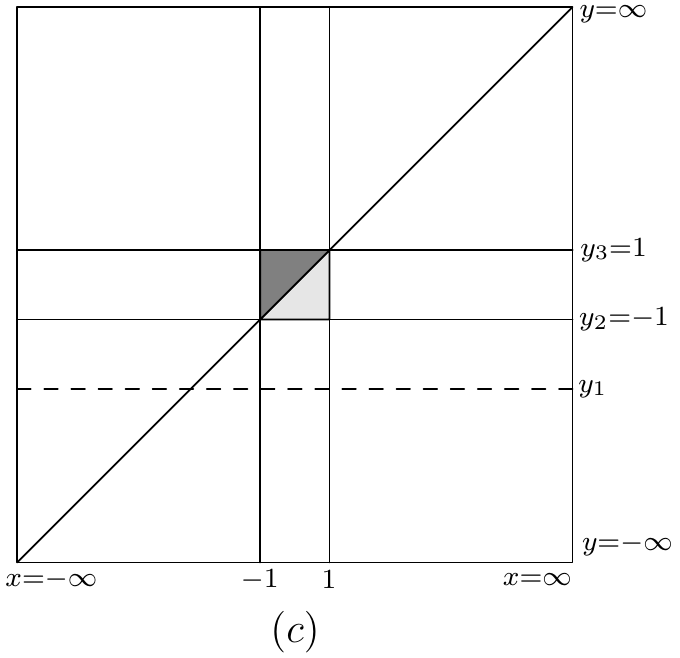}
\end{center}
\begin{center}
\includegraphics[width=.32\textwidth]{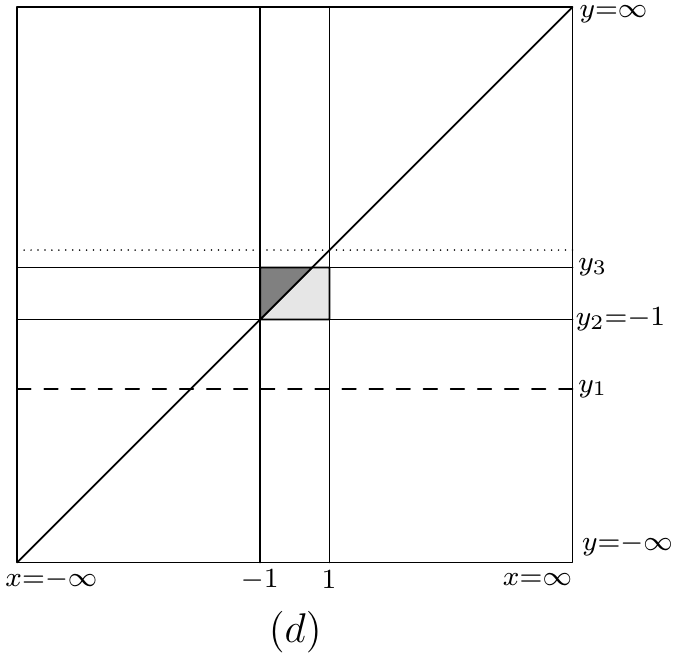}
\includegraphics[width=.32\textwidth]{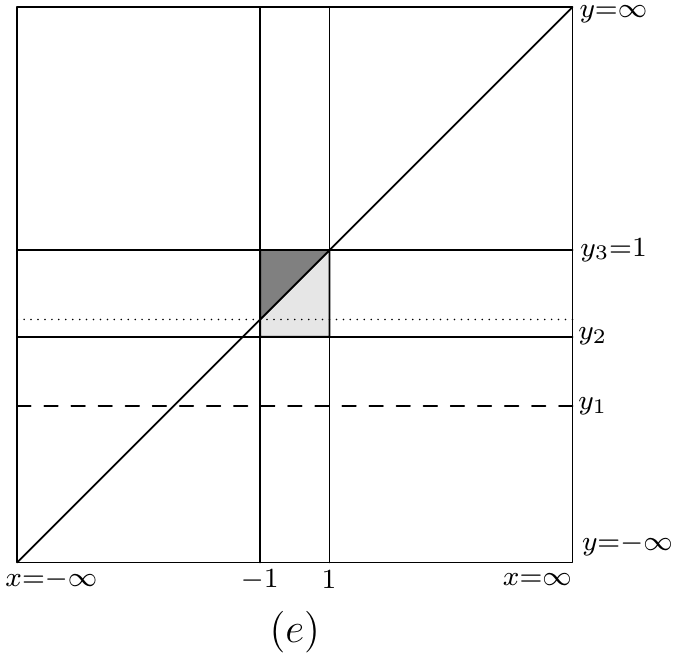}
\includegraphics[width=.32\textwidth]{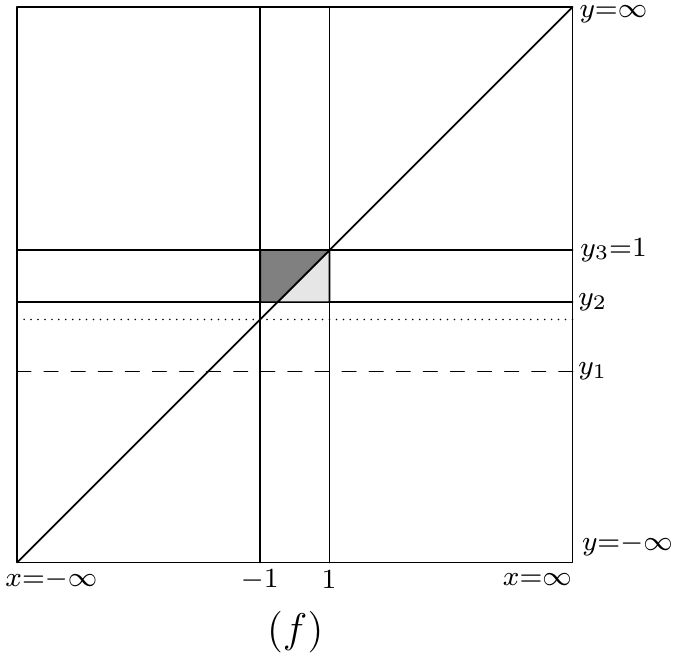}
\end{center}
\caption{Some of the limiting cases.
($a$) $e_2=0$; ($b$) $e_2=-2$ and $\omega^{-1}>3$;
($c$) $e_2=-2$ and $\omega^{-1}=3$; ($d$) $e_2=-2$ and $\omega^{-1}<3$;
($e$) $e_2<-2$ and $\omega^{-1}=1-e_2$;
($f$) $-2<e_2<0$ and $\omega^{-1}=1-e_2$.}
\label{fig2}
\end{figure}

The metric \eqref{ansatz} depends explicitly only on two of the four spacetime coordinates
$(x,y)$.
Therefore let us first try to determine the physical regions for these two coordinates.
Recall that the correct Lorentz signature of the metric \eqref{ansatz} requires $G(x)>0$. This
is achieved by setting $-1<x<1$, or reparametrized as
\[
x=\cos\theta, \quad (0<\theta<\pi).
\]
The alternative choice $x<-\omega^{-1}$ can also yield correct Lorentz signature, but this
choice corresponds to an unbounded $x$ coordinate, which is physically uninteresting because
$x$ is to be interpreted as one of the angular coordinates.
On the other hand, the existence of a static region in the
spacetime requires $F(y)>0$. Rewriting $F(y)$ as
\begin{align}
F(y)=-\omega(y-y_1)(y-y_2)(y-y_3),\label{Fyfactor}
\end{align}
it is clear that the condition $F(y)>0$ requires either $y<y_1$ or $y_2<y<y_3$.
It will become clear shortly that the third root $y=y_3$ of $F(y)$ corresponds to a black hole
event horizon. Therefore, the first region $y<y_1$ is physically relatively less interesting
because the black hole event horizon is out of reach from this region. The second region $y_2<y<y_3$
is more interesting following the same consideration. So, for now, let us assume that $y$
takes values in the second region. Finally, due to the overall conformal factor, we know
that $x=y$ is the conformal infinity of our spacetime, thus the physical region of the
spacetime requires either $y>x$ or $y<x$ but not across $y=x$. In the following, we shall be
concentrating exclusively on the physically interesting region $y>x$. The region $y<x$ may also
be of some physical interests, but it is out of our main focus in this paper.
Depending on the values of $e_2$ and $\omega$, the static region of the spacetime
is bounded by the lines $x=\pm1$, $y=y_2$, $y=y_3$ and the condition $y>x$.
The details are shown in Figs.\ref{fig1} and \ref{fig2}.

In Fig.\ref{fig1}, all physically interesting static regions are shaded in darkgray.
Some of the pictures in Fig.\ref{fig1} also contain a region shaded in lightgray,
which are considered to be unphysical, because they appear on the other side of
conformal infinity. The region shaded in darkgray in Fig.\ref{fig1}($a$) corresponds to
the static region outside a black hole. The lower boundary at $y=y_2$ corresponds to a
compact acceleration horizon as will become clear later. This case is similar to the
de Sitter C-metric in Einstein gravity (cf. Fig.2(a) in ref.\cite{Teo1}) except that the
present spacetime has a non-constant
scalar curvature. The conformal infinity $y=x$ lies behind the
acceleration horizon and is out of reach from this region. The region in darkgray in
Fig.\ref{fig1}($b$) is again a static region outside a black
hole, with lower boundary at $y=y_2$ corresponding to an acceleration horizon. The difference
from the case of Fig.\ref{fig1}($a$) lies in that the acceleration horizon hits the conformal
infinity at some intermediate value $-1<x<1$, so the acceleration horizon cannot be compact
in this case. In Fig.\ref{fig1}($c$), both the upper and lower boundaries of the region shaded in
darkgray hits the conformal infinity at some intermediate value of $x$, so this static region
contains a non-compact black hole event horizon and a
non-compact acceleration horizon. Unlike the case of Fig.\ref{fig1}($c$), the case of
Fig.\ref{fig1}($d$) corresponds to a larger $\omega^{-1}$, which makes the upper shaded region
to have a compact black hole event horizon but the acceleration horizon lies completely
behind the conformal infinity and hence is out of reach from this region.
The upper shaded region in Fig.\ref{fig1}($e$) has a non-compact
black hole event horizon as its upper boundary, which hits conformal infinity at some
$-1<x<1$. The acceleration horizon is beyond reach because it hides behind the conformal
infinity. The last figure, Fig.\ref{fig1}($f$), contains no region shaded in darkgray, which
means that there is no physically interesting static region for this set of parameters.

Fig.\ref{fig1} described only the cases for generic values of $e_2$ and $\omega^{-1}$.
However, there are also some particular values of $e_2$ and $\omega^{-1}$ which
are not shown, i.e. the cases $e_2=0, e_2=-2$ and/or $\omega^{-1}=1-e_2$. These special cases
can be viewed as certain limiting cases of the plots given in Fig.\ref{fig1}. Some of these
limiting cases are depicted in Fig.\ref{fig2}. The limiting case of Fig.\ref{fig1}($f$) at
$y_3\to -1$ is not displayed here because this case is physically uninteresting just like
Fig.\ref{fig1}($f$). In all cases displayed in Fig.\ref{fig2}, either the black hole event
horizon or the acceleration horizon hits the conformal infinity at a single point.

In all cases displayed in Figs.\ref{fig1} and \ref{fig2} except Fig.\ref{fig1}($f$),
the region bounded by the lines $x=\pm1$, $y=y_3$,
$y=+\infty$ and possibly $y=x$ represents the black hole interior which are non-static.
The regions bounded by $x=-1$, $y=y_2$ and $y=x$ in Figs.\ref{fig1}(b), \ref{fig1}(c),
\ref{fig2}(a) and \ref{fig2}(f) are non-static and non-compact which all hide behind the
acceleration horizon at $y=y_2$. Moreover, in Fig.\ref{fig1}(a), the region bounded
by the lines $x=\pm1$, $y=y_1$, $y=y_2$ and $y=x$ corresponds also to a non-static and
non-compact region, and there is yet another static region bounded by $x=-1$, $y=y_1$ and
$y=x$. Here $y=y_1$ may be interpreted as a cosmological horizon (this interpretation of the
acceleration horizon is similar to the case of AdS C-metric in Einstein gravity as shown
in \cite{DisaLemos2002, Kr}).

The descriptions made in the last few paragraphs rely heavily on the statements that the root
$y=y_3$ corresponds to a black hole event horizon and $y=y_2$ corresponds to an acceleration
horizon. Now it is time to justify these statements. For this purpose let us first remark that
the curvature invariants such as $R$, $R_{\mu\nu}R^{\mu\nu}$ and $R_{\mu\nu\rho\sigma}
R^{\mu\nu\rho\sigma}$ are all finite at $y=y_i (i=1,2,3)$, though the concrete expressions
for the latter two curvature invariants are
too lengthy and hence do not worth to be displayed here. This indicates that the roots
of $F(y)$ are coordinate singularities, not essential ones. The true curvature singularity
arises at $y=\pm\infty$, which can already be seen from the expression of Ricci scalar given
in the last section.

To understand the role of the roots of $F(y)$, let us calculate the proper acceleration
of a static observer in the spacetime. The static observer is chosen as
\begin{align}
x^\mu=\left(\frac{A(x-y)}{\sqrt{F(y)}}\eta, y, x, \sigma\right),
\end{align}
where $F(y)$ is given in \eqref{Fy} with $\lambda=0$, $\eta$ is the proper time, and the
spatial coordinates $(y,x,\sigma)$ are fixed to some constants. The proper velocity of this
static observer is
\begin{align}
v^\mu&=\left(\frac{A(x-y)}{\sqrt{F(y)}}, 0, 0, 0\right),
\end{align}
and hence the corresponding proper acceleration reads
\begin{align}
&a^\mu=v^\nu\nabla_\nu v^\mu = A^2(x-y)\,(0,\beta,\gamma,0),
\end{align}
where
\begin{align*}
\beta&=-1+xy-e_2(x+y-e_2)+Am(x+y+y^3-3xy^2)\nonumber\\
&\quad +Ame_2(-2+6xy-3xe_2-3ye_2+2e_2^2),\\
\gamma&=-G(x).
\end{align*}
Therefore, the norm of the proper acceleration is
\begin{align}
|a|^2 =g_{\mu\nu} a^\mu a^\nu=A^2\left(\frac{ \beta^2}{F(y)} +G(x)\right). \label{acc2}
\end{align}
It is important to note that $\beta$ does not have any common factor with $F(y)$, hence
$|a|^2 $ diverges at all three roots of $F(y)$. In other words, all roots of $F(y)$ are
acceleration horizons if accessible from the static region. Among these, $y=y_1$ is not
accessible from the shaded static regions depicted in Figs.\ref{fig1} and \ref{fig2},
so we are left with only
two acceleration horizons at $y=y_2$ and $y=y_3$. In the following, we shall make it clear that
the root $y=y_3$ is actually a black hole event horizon, so only $y=y_2$ is a pure acceleration
horizon.

It should be remarked that the above analysis for the curvature singularity and for
the acceleration horizons applies not only to the
$\lambda=0$ case. For $\lambda>0$, the same analysis still works and the conclusion is unchanged.
The only difference lies in that the roots of $F(y)$ may be different and that the expression for
$\beta$ becomes more complicated.

Since the causal character of conformal infinity is an interesting issue, we try to
make a clarification on this point.
The character of conformal infinity can be studied through the norm of tangent vector of the hypersurface. The tangent vector of conformal infinity is $\Omega_{,\mu}$, where $\Omega=A(x-y)$ and $\Omega=0$ corresponds to conformal infinity. Since $\Omega=0$ corresponds to singularity of the spacetime (\ref{ansatz}), we should calculate the norm of $\Omega_{,\mu}$ in a new spacetime $\mathrm{d}\tilde{s}^2=\Omega^2\mathrm{d}s^2$, where
\begin{align}
\mathrm{d}\tilde{s}^2=-F(y)\mathrm{d}t^2+\frac{\mathrm{d}y^2}{F(y)}
+\frac{\mathrm{d}x^2}{G(x)}+G(x)\mathrm{d}\sigma^2,
\end{align}
with $F(y)$ and $G(x)$ given in (\ref{Gx}) and (\ref{Fy}). Now conformal infinity in spacetime $\mathrm{d}s^2$ corresponds to $\Omega=0$ hypersurface in spacetime $\mathrm{d}\tilde{s}^2$. The norm of $\Omega_{,\mu}$ is
\begin{align}
\Omega_{,\mu}\Omega_{,\nu}\tilde{g}^{\mu\nu}=&- 4 A^3 m x^3+
 6 A^3 e_2 m x^2+(-2 A^2 e_2 + 4 A^3 m - 6 A^3 e_2^2 m) x \nonumber\\
 &+A^2 e_2^2 - 2 A^3 e_2 m +
 2 A^3 e_2^3 m   + 2 e_1^2 \lambda.\label{norm}
\end{align}
Note that in Einstein gravity norm of the tangent vector of conformal infinity depends only on cosmological constant, i.e., the sign of cosmological constant determines the causal character of conformal infinity uniquely. However, for the C-metric of CGCCS model, norm of the tangent vector of conformal infinity depends on the coordinate $x$. In order to determine the sign of the norm (\ref{norm}), we need to find the real roots of the equation $\Omega_{,\mu}\Omega_{,\nu}\tilde{g}^{\mu\nu}=0$, the character of conformal infinity is different on the two sides of the real roots. Solving the equation $\Omega_{,\mu}\Omega_{,\nu}\tilde{g}^{\mu\nu}=0$ roughly we have the following three roots
\begin{align}
x_1&=\frac{e_2}{2}+\frac{X}{6mA^3 2^{2/3}(-Y+\sqrt{4X^3+Y^2})^{1/3}}-\frac{(-Y+\sqrt{4X^3+Y^2})^{1/3}}{12mA^32^{1/3}},\nonumber\\
x_2&=\frac{e_2}{2}-\frac{(1+i\sqrt{3})X}{12mA^3 2^{2/3}(-Y+\sqrt{4X^3+Y^2})^{1/3}}+\frac{(1-i\sqrt{3})(-Y+\sqrt{4X^3+Y^2})^{1/3}}{24mA^3 2^{1/3}},\nonumber\\
x_3&=\frac{e_2}{2}-\frac{(1-i\sqrt{3})X}{12mA^3 2^{2/3}(-Y+\sqrt{4X^3+Y^2})^{1/3}}+\frac{(1+i\sqrt{3})(-Y+\sqrt{4X^3+Y^2})^{1/3}}{24mA^3 2^{1/3}},
\end{align}
with $X = 24m A^5 e_2 - 48m^2 A^6+ 36 m^2 A^6 e_2^2$ and $Y = 864 m^2A^6 e_1^2 \lambda$.
To find which root is real, we need to consider the sign of $4X^3+Y^2$. When $4X^3+Y^2\geq0$,
it is easy to see that $x_1$ is real. If one further requires either $x_2$ or $x_3$ is real,
this would end up with the condition $4X^3+Y^2=0$. Thus for $4X^3+Y^2=0$ we have two real
roots ($x_2$ equals to $x_3$ in this case), and for $4X^3+Y^2>0$ we have only one real root. For
$4X^3+Y^2=0$ the two real roots are
\begin{align}
x_1=\frac{e_2}{2}+\frac{(Y/2)^{1/3}}{6mA^3},\;\;\;\;\;\;x_2
=\frac{e_2}{2}-\frac{(Y/2)^{1/3}}{12mA^3}.
\end{align}
When $4X^3+Y^2<0$, one notes that $X$ is negative and the modulus of
$-Y+\sqrt{4X^3+Y^2}$ is $2|X|^{3/2}$, thus we may write
\begin{align}
-Y+\sqrt{4X^3+Y^2}=2|X|^{3/2}e^{i\theta},
\end{align}
and  $(-Y+\sqrt{4X^3+Y^2})^{1/3}=2^{1/3}|X|^{1/2}e^{i(\theta+2n\pi)/3}$,
where $\theta$ is dependent on the values of $X$ and $Y$ and hence also on $e_1$ and $e_2$.
After a little calculation, one will find that $x_2$ and $x_3$ are real and equal to each other,
i.e.,
\begin{align}
x_2=x_3=\frac{e_2}{2}+\frac{|X|^{1/2}\cos[(\pi+\theta+2n\pi)/3]}{6mA^3}.
\end{align}
$x_1$ is real too, which can also be written as
\begin{align}
x_1=\frac{e_2}{2}-\frac{|X|^{1/2}\cos[(\theta-2n\pi)/3]}{6mA^3}.
\end{align}
In any case, the position of the real roots of the right hand side of \eqref{norm}
is dependent on the values
of $e_1$ and $e_2$, and provided the real roots are sitting in between $x=\pm 1$, the
character of the conformal infinity will change (either from timelike to spacelike or
vice versa) on the two sides of the single root. For double roots the character of the conformal
infinity on both sides are the same, and it changes into lightlike at the roots.

Now let us come back to the $\lambda=0$ case and introduce the following
coordinate transformations
\begin{align}
x=\cos\theta,\quad y-e_2=\frac{1}{A r},\quad t=A \tau.
\label{coordinatestransf}
\end{align}
In terms of these new coordinates, the metric (\ref{ansatz}) can be written as
\begin{align}
\mathrm{d}s^2=\frac{1}{[1-A r (\cos\theta-e_2)]^2}
\left[-Q(r)\,\mathrm{d}\tau^2 +\frac{\mathrm{d}r^2}{Q(r)}
+r^2\left(\frac{\mathrm{d}\theta^2}{P(\theta)}
+P(\theta)\sin^2\theta \,\mathrm{d} \sigma^2\right)\right],
\label{spherical}
\end{align}
where
\begin{align}
Q(r)=(1-A^2 r^2)\left(1-\frac{2m}{r}\right),\quad P(\theta)=1+2m A \cos\theta.\label{PQ}
\end{align}
When $A=0$, the metric \eqref{spherical} degenerates into that of the Schwarzschild black hole,
with $r=2m$ being the event horizon. Note that the root $r=2m$ of $Q(r)$ corresponds to the
root $y=y_3$ of $F(y)$, so $y=y_3$ needs to be the black hole event horizon even at $A\neq 0$.

To interpret the physical meaning of the constant $A$, let us consider the weak field limit
$m\to 0$. In this limit, the second factor in \eqref{acc2} becomes a constant which is
independent of $A$. Thus, {\em the proper acceleration of any static observer in the weak field
limit is proportional to $A$.}

It remains to determine the ranges of the coordinates $t,\sigma$. As a timelike coordinate,
$t$ is unconstrained in the static region, so $-\infty<t<+\infty$. As for the angular
coordinate $\sigma$, let us examine the possible deficit angles around the $\theta=0$ and
$\theta=\pi$ half axes of the $t=const$ and $r=const$ hypersurface. Assuming
$\sigma\in [-\pi C, \pi C]$, the circumference to radius ratios for the infinitesimal circles
around the above two half axes are respectively
\begin{align}
&\lim_{\theta\rightarrow0}\frac{2\pi C P(\theta)\sin\theta}{\theta}=2\pi C(1+2mA),\qquad
(\theta=0),\\
&\lim_{\theta\rightarrow\pi}\frac{2\pi C P(\theta)\sin\theta}{\pi-\theta}=2\pi C(1-2mA),\qquad
(\theta=\pi).\label{conical}
\end{align}
Thus, there exist different conical singularities for the $\theta=0$ and $\theta=\pi$ half axes.
The conical singularity at the $\theta=0$ pole can be canceled by taking $C=1/(1+2mA)$. Then the
deficit angle at $\theta=\pi$ is $8\pi Am/(1+2mA)$. The deficit angle at the $\theta=\pi$ pole
is interpreted as a semi-infinite cosmic string along the $\theta=\pi$ half axes which drags
and accelerates a Schwarzschild-like black hole along the axis \cite{GriffithsPodolsky}.

\subsection{The case $\lambda>0$ with $e_2=0$   \label{section2.2}}

The situation for $\lambda> 0$ is much more complicated than the case $\lambda=0$,
because $F(y)$ is no longer explicitly factorized. To make things easier, let us first
consider the degenerated case $e_2=0$.

When $\lambda>0$ and $e_2=0$, eqs.\eqref{Gx}-\eqref{FG} can be rewritten as
\begin{align}
& G(x) =(1-x)(1+x)(1+\omega x), \label{Gx2}\\
& F(y) = -(1+y)(1-y)(1-\omega y)+\frac{2e_1^2\lambda}{A^2}.\label{Fy2}
\end{align}
In this case, it will be convenient to introduce a coordinate reflection $y\to \tilde y=-y$,
after which the metric becomes
\begin{align}
\mathrm{d}s^2=\frac{1}{A^2(x+\tilde y)^2}\left(\mathcal{F}(\tilde y)\mathrm{d}t^2
-\frac{\mathrm{d}\tilde y^2}{\mathcal{F}(\tilde y)}+\frac{\mathrm{d}x^2}{\mathcal{G}(x)}
+\mathcal{G}(x)\mathrm{d}\sigma^2\right),\label{factorized2}
\end{align}
where
\begin{align}
&\mathcal{G}(\xi)=(1-\xi)(1+\xi)(1+\omega \xi),\label{G2}\\
&\mathcal{F}(\xi)=(1-\xi)(1+\xi)(1+\omega \xi)
-X,\quad \label{F2}\\
&X\equiv \frac{2e_1^2\lambda}{A^2}>0,\nonumber
\end{align}
and clearly
\begin{align}
\mathcal{F}(\xi)=\mathcal{G}(\xi)-X.
\label{FG2}
\end{align}

According to \eqref{FG2}, $\mathcal{F}(\xi)$ differs from $\mathcal{G}(\xi)$ by a negative
constant shift. Meanwhile, $\mathcal{G}(\xi)$ has three explicit real roots thanks to
\eqref{G2}. Since the function $\mathcal{F}(\xi)$ is a polynomial of degree $3$ in $\xi$,
it may have one, two or three real roots depending on the amount of shift, $X$.
We will be mostly interested in the case when $\mathcal{F}(\xi)$ has three real roots,
because this is the case with as many horizons as possible. The other two cases will be
discussed at the end of this subsection.

In order to determine the condition for $\mathcal{F}(\xi)$ to have three real roots, we first
identify its minimum and maximum, which read
\begin{align*}
&\mathcal{F}_{\mathrm{min}}=-\frac{\sqrt{12 \omega ^2+1}+6 \omega ^2 \left(2 \sqrt{12 \omega ^2+1}
+9 X-6\right)+1}{54 \omega ^2},\\
&\mathcal{F}_{\mathrm{max}}=\frac{\sqrt{12 \omega ^2+1}+6 \omega ^2 \left(2 \sqrt{12 \omega ^2+1}
-9 X+6\right)-1}{54 \omega ^2}.
\end{align*}
$\mathcal{F}(\xi)$ will have three real roots if and only if $\mathcal{F}_{\mathrm{min}}<0$
and $\mathcal{F}_{\mathrm{max}}>0$. Recalling that $X>0$, we find that in order
for $\mathcal{F}(\xi)$ to have three real roots, the value of $X$ must be constrained to
be within the range
\begin{align*}
X\in \left(0, \frac{12 \left(\sqrt{12 \omega ^2+1}+3\right) \omega ^2
+\sqrt{12 \omega^2+1}-1}{54 \omega ^2}\right).
\end{align*}
Now assuming that the above condition is fulfilled. Then $\mathcal{F}(\xi)$ can be written
in a factorized form\footnote{Notice that the parameters $a, b, c$ used here and below has
nothing to do
with those appeared in \eqref{abctrans}.}
\begin{align}
{\mathcal{F}}(\xi)&=(\xi-a)(\xi-b)(k_0+k_1\xi).\label{completefactorF}
\end{align}
Similarly we rewrite $\mathcal{G}(x)$ as
\begin{align}
\mathcal{G}(\xi)=(\xi-1)(\xi+1)(p_0+p_1 \xi), \label{completefactorG}
\end{align}
where of course $p_0=-1, p_1=-\omega$.
Comparing eqs.\eqref{completefactorF} and \eqref{completefactorG} with \eqref{G2} and
\eqref{F2}, we get
\begin{align}
&p_0=\Xi\left[-(a+b)^2+ab+1\right]=-1,
\quad
p_1=\Xi(a+b)=-\omega,\nonumber\\
&k_0=\Xi(1+ab),
\qquad\qquad\qquad\qquad\quad\; k_1=\Xi(a+b),
\end{align}
where
\begin{align}
\Xi=\frac{2\lambda e_1^2}{A^2(a^2-1)(1-b^2)}=\frac{X}{(a^2-1)(1-b^2)}. \label{Xi}
\end{align}
Note that we can in principle determine $a,b$ in terms of $\lambda, A, e_1,\omega$ using the
above relations. Now since $p_0, p_1, k_0, k_1$ have a common factor $\Xi$,
we can pick out this common factor and rewrite the metric as
\begin{align}
\mathrm{d}s^2=\frac{1}{\Xi A^2(x+\tilde y)^2}
\left(\tilde {\mathcal{F}}(\tilde y)\mathrm{d}\tilde t\,^2
-\frac{\mathrm{d}\tilde y^2}{\tilde {\mathcal{F}}(\tilde y)}
+\frac{\mathrm{d}x^2}{\tilde {\mathcal{G}}(x)}
+\tilde {\mathcal{G}}(x)\mathrm{d}\tilde \sigma^2\right),\label{factorized3}
\end{align}
where $\tilde t=\Xi\, t,\, \tilde \sigma=\Xi\,\sigma$
and
\begin{align}
\tilde {\mathcal{G}}(\xi)&=\Xi^{-1}{\mathcal{G}}(\xi)=(\xi-1)(\xi+1)[(a+b)(\xi-a-b)+ab+1],
\label{G3}\\
\tilde {\mathcal{F}}(\xi)&=\Xi^{-1} {\mathcal{F}}(\xi)
=(\xi-a)(\xi-b)[(a+b)\xi+ab+1].\label{F3}
\end{align}

\begin{figure}[h]
\begin{center}
\includegraphics[width=.45\textwidth]{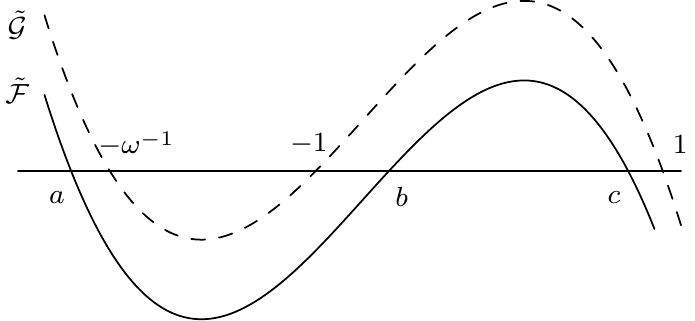}
\end{center}
\caption{$\tilde {\mathcal{F}}(\xi)$ differs from $\tilde {\mathcal{G}}(\xi)$ by a
negative constant shift}
\label{fig3}
\end{figure}

\begin{figure}[h]
\begin{center}
\includegraphics[width=.45\textwidth]{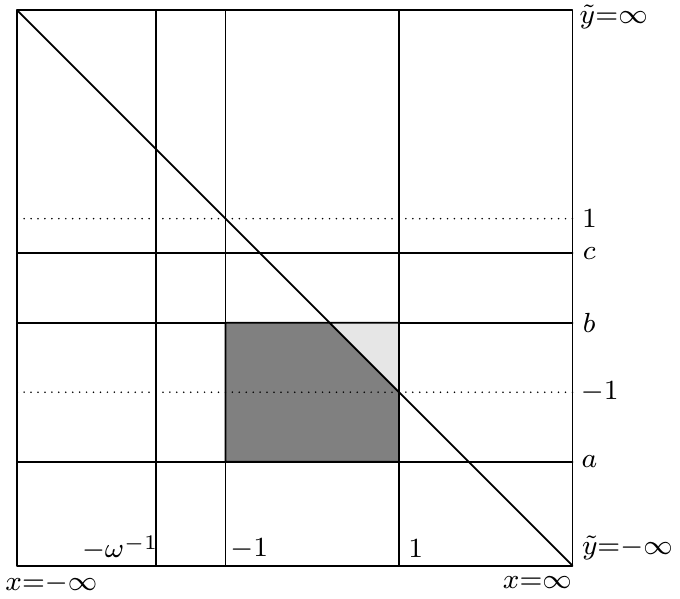}
\end{center}
\caption{Static region for the case $\lambda>0,\, e_2=0$ in the $(x,\tilde y)$ subspace}
\label{fig4}
\end{figure}

The metric (\ref{factorized3}) is invariant under the following two sets of discrete
transformations:
\begin{align}
(1)\quad  &\tilde y\rightarrow-\tilde y, \quad x\rightarrow-x, \quad a\rightarrow-a,
\quad b\rightarrow-b,
\\
(2)\quad & a\leftrightarrow b.
\end{align}
These symmetries allow us to take, without loss of generality, $a+b<0$ and $a<b$. Thus we have
$a<0$. Moreover, it follows from \eqref{FG2} that
\begin{align}
\tilde {\mathcal{G}}(\xi)-\tilde {\mathcal{F}}(\xi)=\Xi^{-1}X=(a^2-1)(1-b^2). \label{FG3}
\end{align}
From \eqref{G3} it is evident that the roots of $\tilde {\mathcal{G}}(x)$ coincides with those
of ${\mathcal{G}}(x)$, i.e. $x=\pm1, -\omega^{-1}$. Just like the $\lambda=0$ case, we restrict
$-1<x<1$, so that $\tilde {\mathcal{G}}(x)>0$. Under this condition, the correct Lorentz
signature of the metric \eqref{factorized3} requires $\Xi>0$ and
$\tilde {\mathcal{F}}(\tilde y)<0$
in the static region. Since we already have $\lambda>0$, the only possibility is to set
$(a^2-1)(1-b^2)>0$ in order that the condition $\Xi>0$ is satisfied.
On the other hand,
from \eqref{F3} we can read off the three real roots of  $\tilde {\mathcal{F}}(\tilde y)$, i.e.
$\tilde y=a,b$ and $\tilde y=c\equiv-\frac{ab+1}{a+b}$. It then follows from \eqref{FG3} that
the roots of $\tilde {\mathcal{G}}(\xi)$ and $\tilde {\mathcal{F}}(\xi)$ are ordered as
\begin{align}
a<-\omega^{-1}<-1<b<c<1, \label{ordroot}
\end{align}
as can be infered from Fig.\ref{fig3}. In the static region, we need to have
$\tilde {\mathcal{F}}(\tilde y)<0$, therefore, this region is bounded by $a<\tilde y <b$
or $\tilde y>c$.
Besides the two pairs $x=\pm1$, $\tilde y=a,b$ of boundaries, the static region is also
constrained by the conformal infinity which is now defined by $x+\tilde y=0$. Assuming
$\tilde y<-x$ (which is equivalent to $y>x$ in the last subsection), we can depict the
static region of the $\lambda>0, e_2=0$ case of our solution as the dark shaded region in
Fig.\ref{fig4}. In this case, the region $-1<x<1, -\infty<\tilde y <a$ represents
the black hole interior. The region bounded by $x=-1, \tilde y =b, \tilde y =c$ and
$\tilde y =-x$ corresponds to a non-static region with $\tilde y =c$ acting as a non-compact
cosmological horizon. When $b=c$, the accelerating horizon and the cosmological horizon
coincide, and the above non-static region cease to exist.
Finally, the region bounded by $x=-1,\tilde y=c$ and $\tilde y=-x$
is a static region from which the black hole event horizon is inaccessible. This last static
region always exists.

At the end of this subsection, let us consider the case when $\mathcal{F}(\xi)$ has only one
real root. In this case, we have $X>\frac{12 \left(\sqrt{12 \omega ^2+1}+3\right) \omega ^2
+\sqrt{12 \omega^2+1}-1}{54 \omega ^2}$, and the static region of the spacetime is bounded
by the lines $x=\pm1$, $\tilde y=a$ and $\tilde y=-x$, and the line $\tilde y=a$ still
corresponds to the black hole event horizon. Diagrammatically this case corresponds to
Fig.\ref{fig4} with the two horizontal lines $\tilde y =b$ and $\tilde y =c$ removed.

\subsection{$\lambda>0$ and $e_2\neq 0$}

Now let us consider the most general case $\lambda>0$ with $e_2\neq 0$. In this case,
after making a coordinate reflection
\[
y\to \tilde y =-y,
\]
and repeating the process that had led to the meric form \eqref{factorized3},
we can rewrite the metric \eqref{ansatz} as
\begin{align}
\mathrm{d}s^2=\frac{1}{\Xi A^2(x+\tilde y)^2}
\left(\tilde {\mathcal{F}}(\tilde y+e_2)\mathrm{d}\tilde t\,^2
-\frac{\mathrm{d}\tilde y^2}{\tilde {\mathcal{F}}(\tilde y+e_2)}
+\frac{\mathrm{d}x^2}{\tilde {\mathcal{G}}(x)}
+\tilde {\mathcal{G}}(x)\mathrm{d}\tilde \sigma^2\right),\label{factorized4}
\end{align}
where $\tilde t=\Xi\, t,\, \tilde \sigma=\Xi\,\sigma$, the constant $\Xi$ is the same as in
\eqref{Xi}, and $\tilde {\mathcal{G}}(\xi)$, $\tilde {\mathcal{F}}(\xi)$ are still given by
eqs.\eqref{G3} and \eqref{F3}. Comparing to the case of $\lambda>0$ with $e_2=0$, the only
difference lies in the coordinate shifts in the metric function $\tilde {\mathcal{F}}(\tilde y)
\to \tilde {\mathcal{F}}(\tilde y+e_2)$.
Consequently the three roots of
$\tilde {\mathcal{F}}(\tilde y+e_2)$ are given by
\[
\tilde y_1=a-e_2, \quad \tilde y_2=b-e_2, \quad \tilde y_3=c-e_2,
\]
i.e. all three roots get shifted by the same amount
$-e_2$. Since the constants $a,b,c$ still obey the constraint \eqref{ordroot}, different
amounts of shift will result in different orders of the roots $\tilde y_1, \tilde y_2,
\tilde y_3$ when compared with the roots $x=\pm1, -\omega^{-1}$ of $\tilde {\mathcal{G}}(x)$.
Consequently, the physically interesting static region of the spacetime bounded by $x=\pm1$ and
$\tilde y=\tilde y_1$, $\tilde y=\tilde y_2$ and possibly the conformal infinity
$\tilde y=-x$ will get shifted upwards or downwards as compared to the case of Fig.\ref{fig4}.
Depending on the values of $e_2$ and $b-a$, there are the following possibilities:
\begin{itemize}
\item $e_2>0$. Such cases are depicted in Fig.\ref{fig5} and the static regions are shaded
in darkgray.
\begin{figure}[h]
\begin{center}
\includegraphics[width=.45\textwidth]{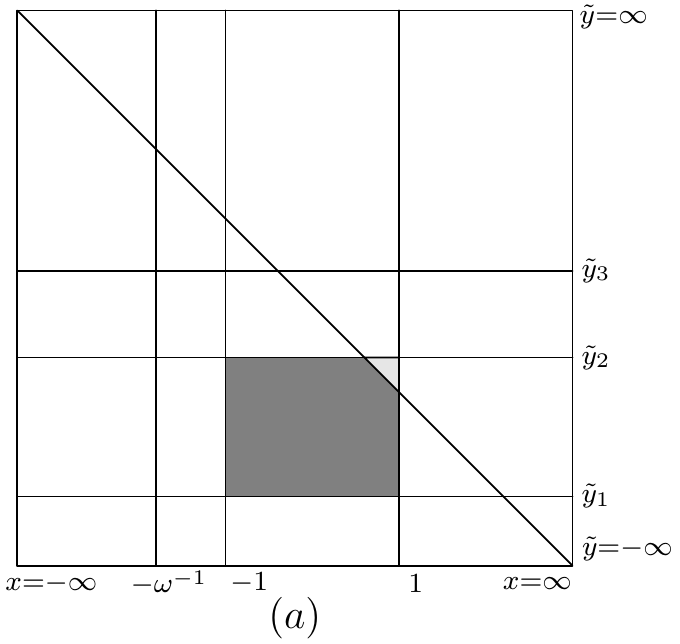}
\includegraphics[width=.45\textwidth]{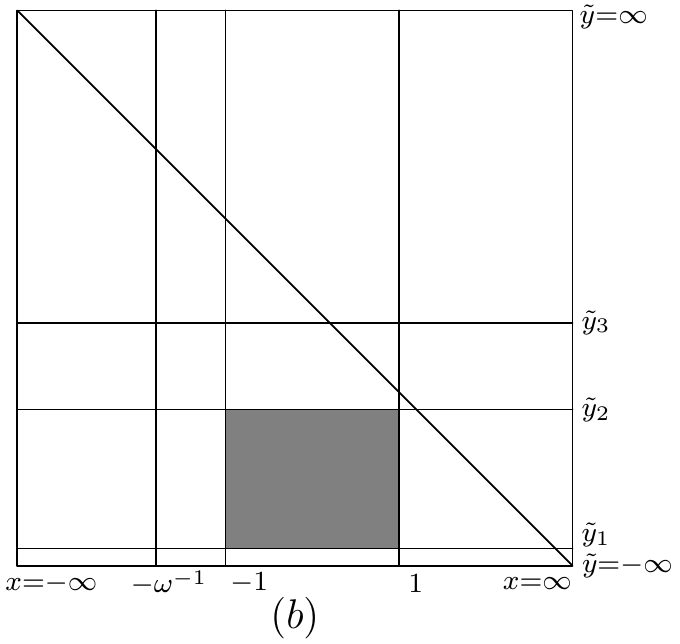}
\end{center}
\caption{Static region for the case $e_2>0$: $(a)$ $0<e_2<1+b$; $(b)$ $e_2>1+b$.}
\label{fig5}
\end{figure}
\item $e_2<0$ and $b-a>2$. Such cases are depicted in Fig.\ref{fig6} and the static regions are
shaded in darkgray.
\begin{figure}[h]
\begin{center}
\includegraphics[width=.23\textwidth]{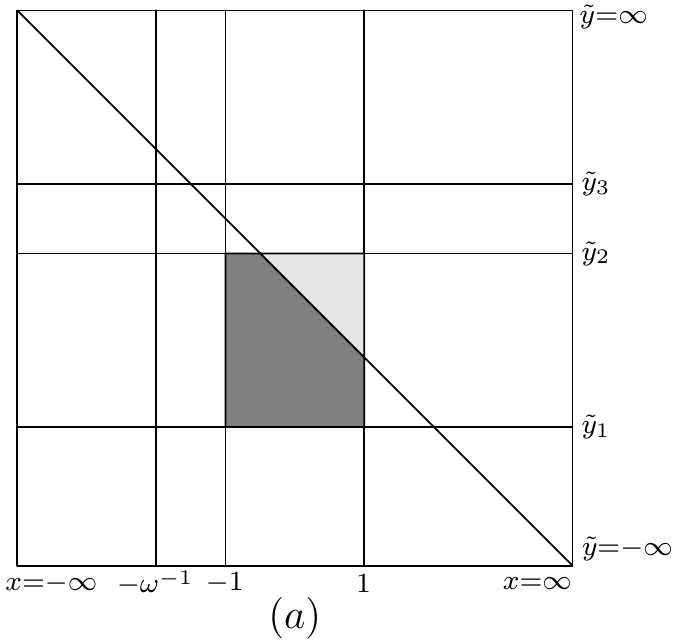}
\includegraphics[width=.23\textwidth]{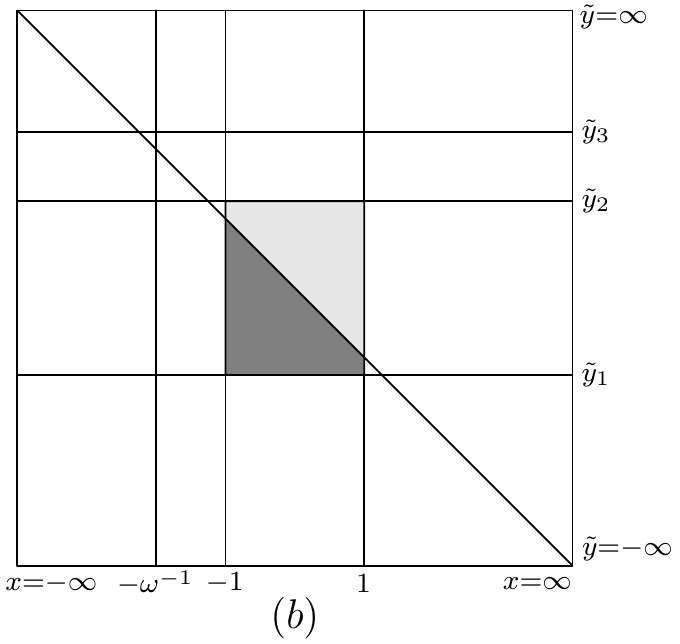}
\includegraphics[width=.23\textwidth]{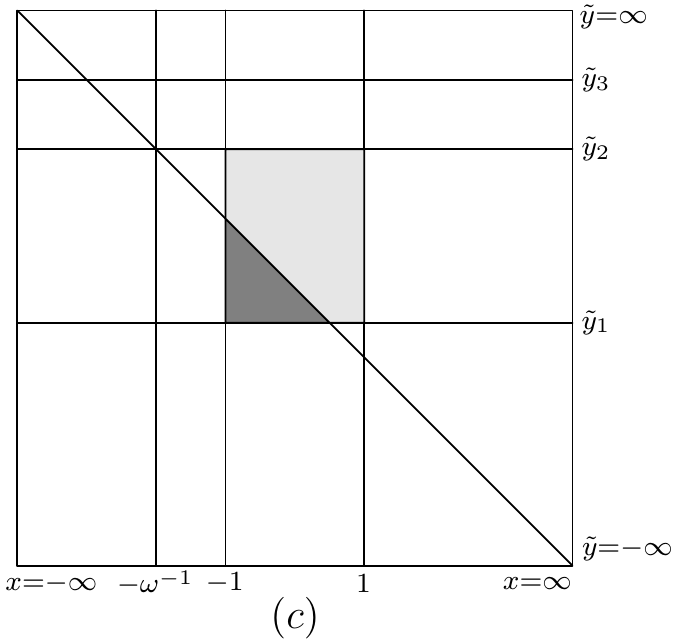}
\includegraphics[width=.23\textwidth]{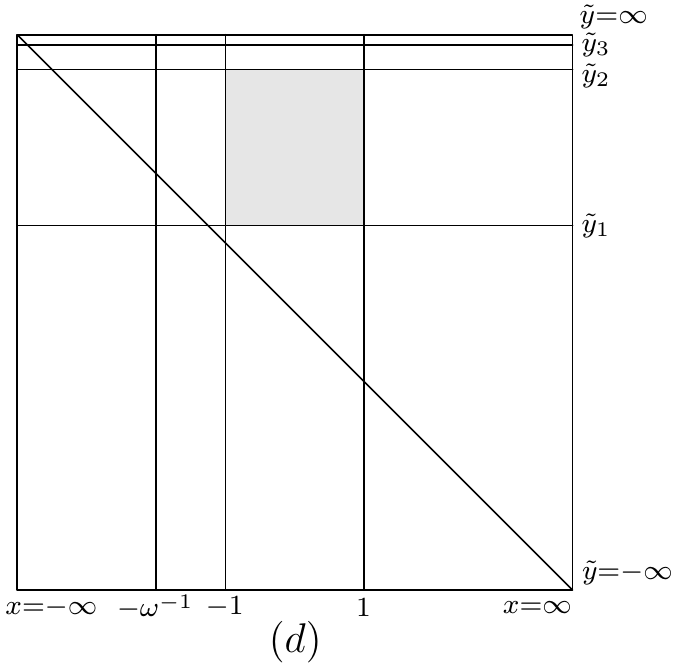}
\end{center}
\caption{Static region for the case $e_2<0$ and $b-a>2$: ($a$) $b-1<e_2<0$;
($b$) $a+1<e_2<b-1$; ($c$) $a-1<e_2<a+1$; ($d$) $e_2<a-1$.}
\label{fig6}
\end{figure}
\item $e_2<0$ and $b-a<2$. Such cases are depicted in Fig.\ref{fig7} and the static regions are
shaded in darkgray.
\begin{figure}[h]
\begin{center}
\includegraphics[width=.23\textwidth]{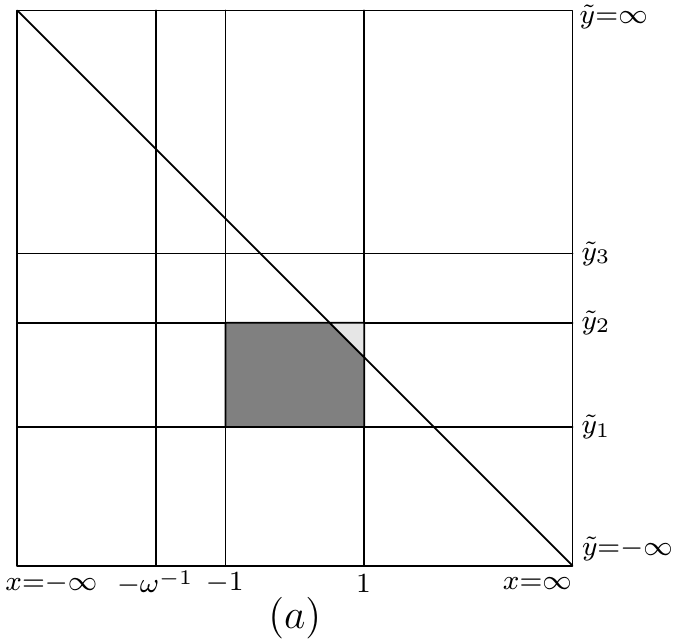}
\includegraphics[width=.23\textwidth]{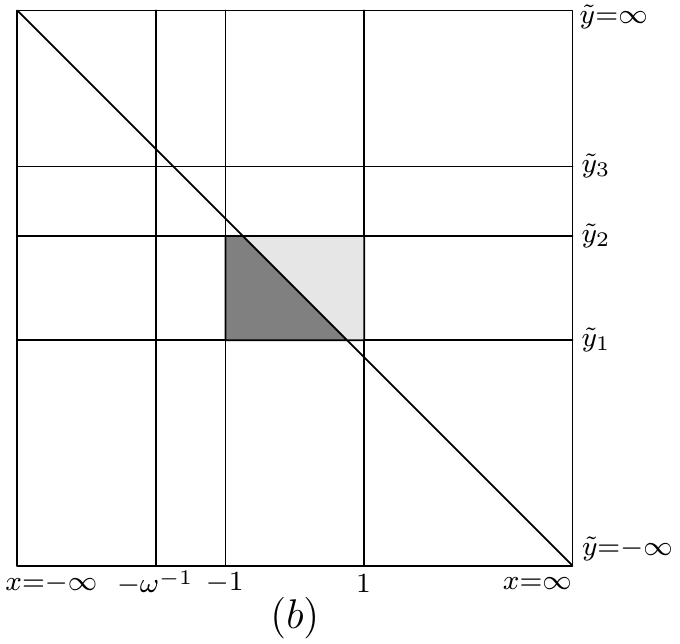}
\includegraphics[width=.23\textwidth]{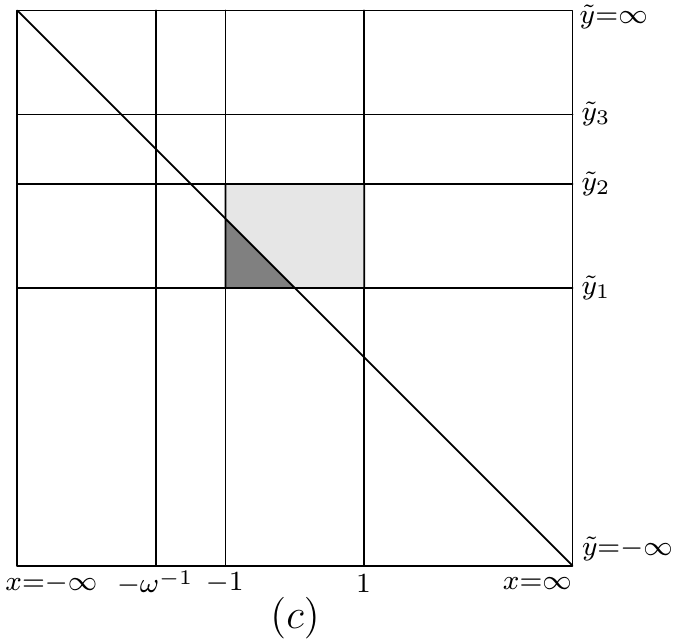}
\includegraphics[width=.23\textwidth]{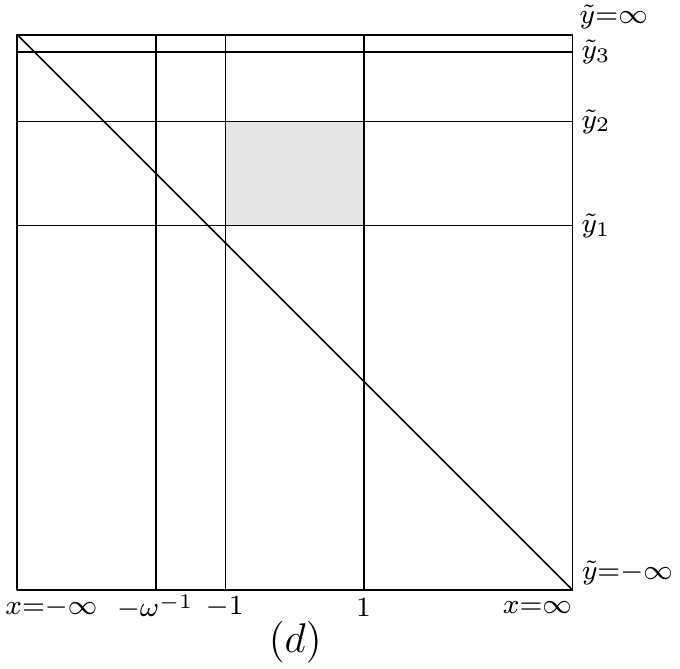}
\end{center}
\caption{Static region for the case $e_2<0$ and $b-a<2$:
($a$) $a+1<e_2$; ($b$) $b-1<e_2<a+1$; ($c$) $a-1<e_2<b-1$; ($d$) $e_2<a-1$.}
\label{fig7}
\end{figure}
\end{itemize}
These figures exhaust all possible physically interesting parameter ranges of our solution
when $\tilde {\mathcal{F}}(\tilde y+e_2)$ has three real roots. We may, of course, discuss the
cases when $\tilde {\mathcal{F}}(\tilde y+e_2)$ has two or one real roots just like we did at
the end of the last subsection. Since these cases are relatively simpler than the case
with three real roots, we omit the corresponding details.

{\em Remarks:}
\begin{enumerate}
\item Fig.\ref{fig6}$(d)$ and Fig.\ref{fig7}$(d)$ do not contain any region shaded in darkgray,
so the parameter ranges corresponding to these two figures are physically uninteresting;
\item In all cases in Figs.\ref{fig5}, \ref{fig6} and \ref{fig7} except
Fig.\ref{fig6}$(d)$ and Fig.\ref{fig7}$(d)$, the lines $\tilde y=\tilde y_1$ correspond
to black hole event horizons;
\item Whenever accessible from the static region, the lines $\tilde y=\tilde y_2$ correspond
to acceleration horizons and the lines $\tilde y=-x$ correspond to conformal infinities;
\item Some limiting cases are not depicted in Figs.\ref{fig5}, \ref{fig6} and \ref{fig7}. Such
limiting cases occur when one or two of the conners of the rectangle bounded by
$x=\pm1$ and $\tilde y=\tilde y_1$, $\tilde y=\tilde y_2$ happen to lie on the line
$\tilde y=-x$. In such cases, either the black hole event horizon or the acceleration horizon
hit the conformal infinity at a single point;
\item In all physically interesting cases, the regions bounded by $x=\pm1$,
$\tilde y=\tilde y_1$, $\tilde y=-\infty$ and possibly $\tilde y=-x$ correspond to the black
hole interior.
\end{enumerate}

\section{Other conformal gauges\label{sec4}}

The analysis for the spacetime properties made in the last section has relied purely on
the explicit form of the metric but not on the action of the CGCCS model. However,
as mentioned in the introduction, an important feature of the model under study is the
conformal invariance under $g_{\mu\nu}\rightarrow \Omega^2(x)g_{\mu\nu},
\Phi\rightarrow\Omega^{-1}(x)\Phi$, which means that if $g_{\mu\nu}$ is a solution, then
$\Omega^2(x)g_{\mu\nu}$ is also a solution for any smooth function $\Omega(x)$. In this
section, we would like to discuss some of the consequences or implications of the
conformal invariance.

First of all, following the discussions made in the last two sections, it seems that the
C-metric solution \eqref{ansatz} with $F(y)$ and $G(x)$ given in \eqref{Fy}, \eqref{Gx}
does not asymptote to an (A)dS spacetime, therefore, it is not clear how to implement
Maldacena's Neumann boundary condition, or whether such a boundary condition can be implemented
in principle in our case. Under such circumstances, we have no other choices but considering
all $\Omega(x)$ on equal footing. However, writing down the explicit form of the metric
implies a concrete, fixed choice for $\Omega(x)$, i.e. we have to work in concrete
conformal gauges. In fact, since the Hawking temperature of a black hole depends only on
the explicit form of the metric but not on the action of the gravity model, any explicit
form of the metric of a black hole spacetime implies a definite Hawking temperature and hence
a prescribed energy scale, which signifies the breaking of conformal symmetry. This is the
situation for all black hole solutions to conformal gravity.

Secondly, unlike the usual gauge symmetries in quantum field theories in which the gauge
choice is irrelevant to the physics, different choices of conformal gauges corresponds to
different physics (or physics at different scales). Another important aspect of conformal
gauge choice is related to the kinematics in curved spacetime. Although the action of the
model under investigation is invariant under conformal transformations, different
choices of conformal gauges yield different Christoffel connections and thus different
geodesics. This fact actually lies behind the reason why conformal gravity can fit the galaxy
rotation curves quite good while Einstein gravity fails to do so without the introduction
of dark matter \cite{Mannheim}.

Last but not the least, let us remark that unlike the situation of Einstein gravity in
which only regular conformal mappings can be applied while constructing the Carter-Penrose
diagrams, in our case {\it any conformal factor} $\Omega^2(x)$ can be applied,
even if it contains some isolated singularities, since such conformal factors
do map one solution of the field equation to another\footnote{When talking about exact
solutions, one always avoids the singularities, otherwise even the Schwarzschild metric
will not solve the standard Einstein equation because the metric simply loses any meaning at the
singularity.}. In fact, the use of singular conformal factor to map one solution of
conformal gravity to another is a usual practice in the literature \cite{Lu:2012xu}.

Among the infinite many choices of conformal gauges,
we are particularly interested in two other gauge choices, which are closely related to the
answer of the following two questions:

Q1. Is our solution contained in the conformal class of an Einstein manifold?

Q2. What are the major differences between all the different conformal gauges?

The answer to Q1 is simply ``yes''. To see this, we would like to make a
conformal transformation, bringing the original metric \eqref{ansatz} into another
conformal gauge, in which the metric becomes that of a constant curvature spacetime.
Concretely, we choose the following Weyl factor $\Omega(x)$,
\begin{align}
\Omega(x)=\frac{x-y}{x+y-e_2}. \label{Om}
\end{align}
Then, after performing the conformal transformation \eqref{conft}, the metric becomes
\begin{align}
\mathrm{d}s^2=\frac{1}{A^2(x+y-e_2)^2}\left(-F(y)\mathrm{d}t^2+\frac{\mathrm{d}y^2}{F(y)}
+\frac{\mathrm{d}x^2}{G(x)}+G(x)\mathrm{d}\sigma^2\right),\label{ansatzs}
\end{align}
where $G(x)$ and $F(y)$ are still given by \eqref{Gx} and \eqref{Fy}. On this occasion it is
tempting to make a shift of coordinate $y\to \bar y=y-e_2$, after which the metric becomes
\begin{align}
\mathrm{d}s^2=\frac{1}{A^2(x+\bar y)^2}\left(-\bar F(\bar y)\mathrm{d}t^2
+\frac{\mathrm{d}\bar y^2}{\bar F(\bar y)}
+\frac{\mathrm{d}x^2}{G(x)}+G(x)\mathrm{d}\sigma^2\right),\label{ansatzs2}
\end{align}
where $\bar F(\bar y)$ is given by
\begin{align}
\bar F(\bar y)=-(1+\bar y)(1-\bar y)(1-2mA\bar y)+\frac{2e_1^2\lambda}{A^2}. \label{bary}
\end{align}
The metric \eqref{ansatzs2} is precisely the AdS C-metric with the cosmological constant
\[
\Lambda=-6e_1^2\lambda,
\]
which appeared first in \cite{PB} and was analyzed in
detail in \cite{DisaLemos2002,Teo1} in the context of standard Einstein gravity. Now we
recovered the same metric in conformal gravity.
The first few curvature invariants of the spacetime \eqref{bary} are
\begin{align*}
R=-24\lambda e_1^2,\quad R_{\mu\nu}R^{\mu\nu}=144\lambda^2e_1^4,\quad
R_{\mu\nu\rho\sigma}R^{\mu\nu\rho\sigma}=48A^6m^2(x+\bar{y})^6+96e_1^4\lambda^2.
\end{align*}
The only curvature singularity is located at $\bar y=\infty$. Notice that the constant $e_2$
disappeared completely in the metric \eqref{ansatzs2}. As a sharp comparison, we have seen
enough on the importance of the constant $e_2$ while exploring the structure of the spacetime
in the original gauge \eqref{ansatz} in the last section. Notice also that after making the
conformal transformation using the Weyl factor \eqref{Om}, the scalar field $\Phi$ in \eqref{Phi}
becomes a constant,
\[
\Phi=e_1,
\]
therefore, the on-shell kinetic energy of the scalar field is zero
and the on-shell scalar potential becomes an effective cosmological constant.
In this case, the on-shell action of the model
under investigation becomes that of the Einstein-Weyl gravity (i.e. $-R+C^2$ gravity) with a
cosmological constant, which is neither that of pure Einstein gravity nor that of pure conformal
gravity. Therefore, finding out that our solution to the CGCCS model lies in the conformal
class of AdS C-metric is a nontrivial fact rather than just another special case that fits in
Maldacena's argument for pure conformal gravity.
Last but not the least, let us remark that the metric \eqref{ansatzs2}
looks extremely similar to the $e_2=0$ case of the original metric \eqref{ansatz}, however
with a big difference in the position of conformal infinity. For \eqref{ansatz}, the conformal
infinity lies at $y=x$, while for \eqref{ansatzs2}, the conformal infinity lies at $\bar y=-x$.
So, if one intends to depict the static region of the spacetime \eqref{ansatzs2},
she/he would have
ended in a diagram like Fig.\ref{fig2}$(a)$, but with the direction of the line representing
conformal infinity changed from SW-NE to NW-SE.

Another conformal gauge which we would like to mention is the gauge
\begin{align}
\mathrm{d}s^2=\frac{1}{A^2y^2}\left(-F(y)\mathrm{d}t^2+\frac{\mathrm{d}y^2}{F(y)}
+\frac{\mathrm{d}x^2}{G(x)}+G(x)\mathrm{d}\sigma^2\right),\label{metricgauge3}
\end{align}
which can be arrived in from \eqref{ansatz} via the transformations
\begin{align}
g_{\mu\nu}\rightarrow \frac{(x-y)^2}{y^2}g_{\mu\nu}, \quad \Phi\rightarrow\frac{y}{x-y}\Phi,
\label{conformaltransf}
\end{align}
after which the scalar field takes the value
\[
\Phi=\frac{e_1 y}{x+y-e_2}.
\]
Notice that unlike the other two
conformal gauges discussed previously, the conformal infinity in the present case appear at
$y=0$. This makes the coordinate $y$ to be reminiscent to the famous Poincare coordinate
for AdS spacetime. However, the metric \eqref{metricgauge3} is not AdS,
and is not even a constant curvature spacetime, as can be easily seen from the Ricci scalar
\[
R=-12{A}^{2} \left\{ mA \left[(e_2-x)y^2+(1-3e_2^2)y+2e_2(e_2^2-1)\right]
-e_2 y+e_2^2-1\right\} -24e_2^2\lambda.
\]
The curvature singularities appear at $|y|\to \infty$. Due to the appearance of the conformal
infinity, we must take either $y\geq0$ or $y\leq0$ when considering the structure of the
spacetime. We take the former choice $y\geq0$.

Assuming that the ratio of parameters $e_1/A$ is taking values in the appropriate range,
we can rewrite $F(y)$ in a completely factorized form
\begin{align}
F(y)=(y-a_1)(y-a_2)(\delta -2m A y), \label{FyF}
\end{align}
where $a_1,a_2,\delta$ are real numbers which are determined implicitly by the
original parameters $e_1,e_2,m,A$. Without loss of generality, let us assume the three
roots $y=a_1,a_2,a_3\equiv \delta/(2mA)$ are ordered as
\[
a_1< a_2< a_3.
\]
Then the correct Lorentz signature of the metric requires $-1<x<1$ and $a_2<y<a_3$.
It can be verified using the same arguments that have led to \eqref{acc2} that the roots
$y=a_2$ and $y=a_3$ are exactly where the proper acceleration of static observers tends to
diverge. With a little bit more efforts we can identify that $y=a_3$ is the black hole event
horizon and $y=a_2$ is a pure acceleration horizon, provided that these roots are
accessible from the physically interesting static region of the spacetime.

\begin{figure}[h]
\begin{center}
\includegraphics[width=.32\textwidth]{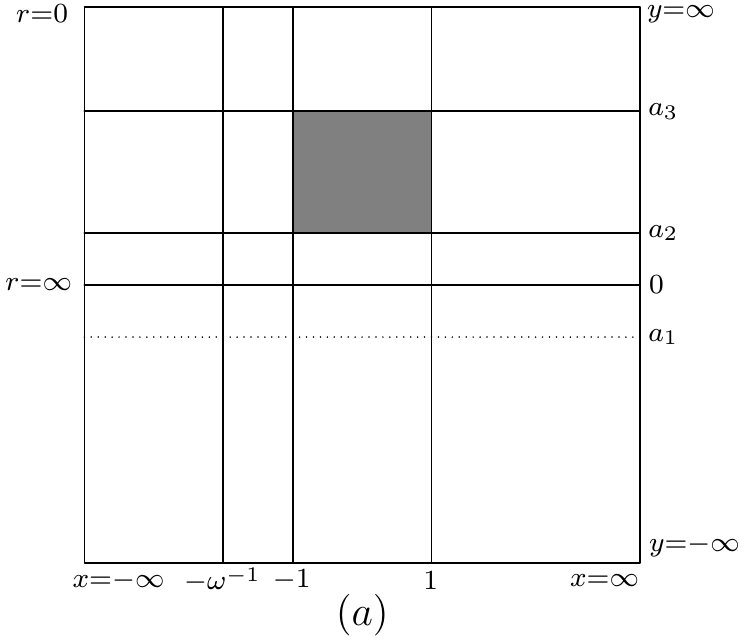}
\includegraphics[width=.32\textwidth]{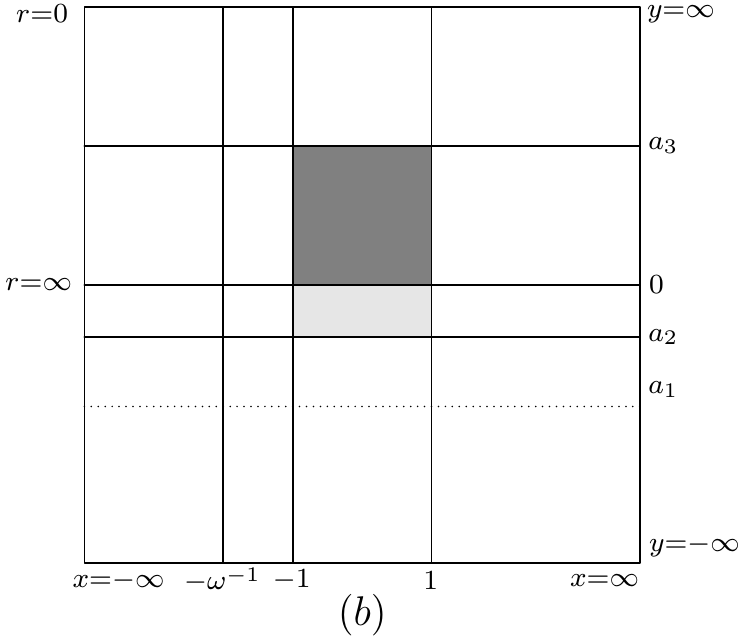}
\includegraphics[width=.32\textwidth]{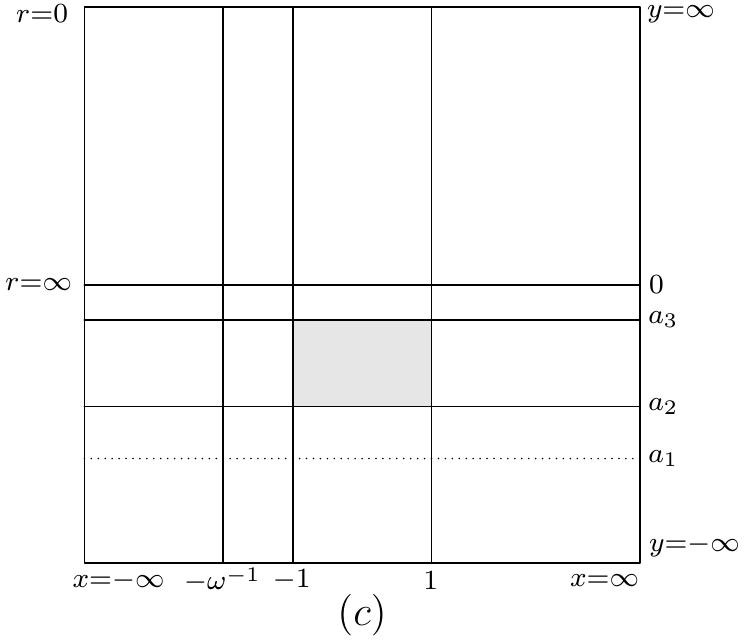}
\end{center}
\caption{Static regions for the metric \eqref{metricgauge3}:
($a$) $0<a_2<a_3$; ($b$) $a_2<0<a_3$;
($c$) $a_2<a_3<0$.}
\label{fig8}
\end{figure}

Depending on the signatures of $a_2$ and $a_3$, there are
three possibilities which are depicted in Fig.\ref{fig8}. In Fig.\ref{fig8}$(a)$,
the physically interesting region (shaded in darkgray) is bounded by $x=\pm1$, a black hole
event horizon at $y=a_3$ and an acceleration horizon at $y=a_2$. The conformal infinity
$y=0$ lies behind the acceleration horizon and hence is out of reach from the static region.
The case of Fig.\ref{fig8}$(b)$ is different in that the conformal infinity appears first and
so the acceleration horizon is beyond reach from the static region.
A possible degeneration of Fig.\ref{fig8}$(a)$ and Fig.\ref{fig8}$(b)$ occurs at $a_2\to 0$,
in which case the accelerating horizon and the conformal infinity coincide.
In Fig.\ref{fig8}$(c)$,
both $y=y_2$ and $y=y_3$ are located on the other side of the conformal infinity and therefore
this case is physically uninteresting.

If we introduce the following coordinate transformations
\begin{align*}
&y \to \frac{1}{Ar},\quad x\to \cos\theta,\quad t\to A\tau,
\end{align*}
the metric \eqref{metricgauge3} can be brought into the following form,
\begin{align}
\mathrm{d}s^2=-\mathcal{Q}(r)\mathrm{d}\tau^2
+\frac{\mathrm{d}r^2}{\mathcal{Q}(r)}
+\frac{r^2\mathrm{d}\theta^2}{\mathcal{P}(\theta)}
+\mathcal{P}(\theta)r^2\sin^2\theta \mathrm{d}\sigma^2,
\label{metricgauge32}
\end{align}
where
\begin{align*}
&\mathcal{Q}(r)=\left(1-a_1 Ar\right)\left(1-a_2 Ar\right)
\left(\delta-\frac{2m}{r}\right),\\
&\mathcal{P}(\theta)=1+2mA \cos\theta.
\end{align*}
In this coordinate system, the overall conformal factor in the metric completely disappeared
and the metric looks like an ordinary static black hole spacetime which bears no resemblance to
the C-metric. However this is only superficial. Actually, \eqref{metricgauge32} still
corresponds to an accelerating black hole spacetime, if the parameters are taken in the
appropriate ranges. As a consequence, it can be static but not spherically symmetric
unless $m=0$.

Now reviewing the three different conformal gauges we have discussed so far, we come to the
following conclusion, which is also part of the answer to Q2, i.e. the location of the
conformal infinities are significantly affected by the choice of conformal gauges, and thus
also the horizon structures are quite different in different conformal gauges.
This is one of the major differences between different choices of conformal gauges.

\section{Concluding remarks\label{section4}}

We have thus presented an exact C-metric solution to the CGCCS model, in which the scalar
field played as a nontrivial matter source. When the parameters are in the appropriate ranges,
the solution may contain a black hole event horizon and an acceleration horizon and both
horizons may be cut by the conformal infinity or be hidden behind the conformal infinity.
The solution belongs to the class of Petrov type D spacetimes and is conformal to the
standard cosmological C-metric known in vacuum Einstein gravity.

For all parameter ranges we studied in detail the physically interesting static regions
as depicted in Figs.\ref{fig1}-\ref{fig8} except Fig.\ref{fig3}.
However these figures do not exhaust all possible
static regions. There are other static regions like the ones shaded in lightgray in the
above mentioned figures in which are not of major concern in this paper. The complete
understanding of the spacetime represented by our solution is still awaiting to be done,
and in particular an analysis on the global structure in the new relative sense
as mentioned in Sec.\ref{sec4} may be a good starting point. Anyway, we have seen plenty
reasons to expect that the C-metric solution to the CGCCS model contain much richer physics
as compared to the
C-metric solution of Einstein gravity.

\section*{Acknowledgment}

KM and LZ thank the anonymous referee for useful comments. This work is supported by the National Natural Science Foundation of China (NSFC) under the
grant numbers 11447153 (for KM) and 11575088 (for LZ).

\providecommand{\href}[2]{#2}\begingroup
\footnotesize\itemsep=0pt
\providecommand{\eprint}[2][]{\href{http://arxiv.org/abs/#2}{arXiv:#2}}

\end{document}